# Imaging of van der Waals Materials via Standing-Wave Photoemission Microscopy: Depth-Resolved Electronic Structure of $WS_2$

Jay R. Paudel[1,2], Ryan Muzzio[3], Matthew E. Matzelle[4,5], Sharup Sheikh[1], Ahmed Fadul[4,5], Aalok Tiwari[3], Farhad Salmassi[6], Eric Gullikson[6], Kathleen M. McCreary[7], Berend T. Jonker[7], Jouko Nieminen[8,4], Claus M. Schneider[9], Florian Kronast[10], Arun Bansil[4,5], Jyoti Katoch[3], and Alexander X. Gray[1]

*[1] Physics Department, Temple University, Philadelphia, Pennsylvania 19122, USA*
*[2] Chemical Sciences Division, Lawrence Berkeley National Laboratory, Berkeley, California 94720, USA*
*[3] Department of Physics, Carnegie Mellon University, Pittsburgh, Pennsylvania 15213, USA*
*[4] Department of Physics, Northeastern University, Boston, Massachusetts 02115, USA*
*[5] Quantum Materials and Sensing Institute, Northeastern University, Burlington, Massachusetts 01803, USA*
*[6] Center for X-ray Optics, Lawrence Berkeley National Laboratory, Berkeley, California 94720, USA*
*[7] U.S. Naval Research Laboratory, Washington, District of Columbia 20375, USA*
*[8] Computational Physics Laboratory, Tampere University, Tampere 33014, Finland*
*[9] Peter Grünberg Institut (PGI-6), Forschungszentrum Jülich GmbH, D-52425 Jülich, Germany*
*[10] Helmholtz-Zentrum Berlin für Materialien und Energie, 12489 Berlin, Germany*

**email: axgray@temple.edu*

Two-dimensional van der Waals materials promise electronic, optoelectronic, and quantum technologies, yet depth-resolved characterization remains challenging. Here, we demonstrate standing-wave photoemission electron microscopy (SW-PEEM) for Å-scale spectromicroscopy of monolayer $WS_2$ on a W/C multilayer substrate. Tuning the X-ray standing wave through the monolayer yields a chemical depth profile and valence-band modulation with enhanced sensitivity to the top and bottom sulfur layers. X-ray optical modeling determines the structure and field distribution. The measurements reveal an ~0.2 eV shift in sulfur-derived valence-band spectral weight between measurements with enhanced sensitivity to the top and bottom sulfur layers. This shift is unlikely to arise from strong direct substrate hybridization and is instead consistent with sulfur-related surface species, as supported by calculations using a representative elemental-sulfur model. These results establish SW-PEEM as a non-destructive depth-resolved probe, highlighting its potential to probe interfacial coupling, chemical reconstruction, and emergent states in van der Waals and moiré systems.

Two-dimensional (2D) van der Waals materials and their heterostructures have emerged as versatile platforms for next-generation electronic, optoelectronic, and quantum technologies [1-5]. Their atomically thin nature, combined with strong electronic correlations and interlayer coupling effects, makes them highly sensitive to external perturbations, enabling novel functionalities in ultrathin device architectures [6-10]. Despite significant progress in their synthesis and characterization, a key challenge remains: achieving a comprehensive understanding of their electronic and chemical properties in three dimensions. Most existing scanning spectromicroscopic techniques, such as scanning tunneling microscopy (STM), scanning near-field optical microscopy (SNOM), and nano-angle-resolved photoemission spectroscopy (nano-ARPES), offer high spatial resolution but are inherently two-dimensional, limiting their ability to resolve depth-dependent variations in electronic structure. A method capable of directly probing the depth- and layer-resolved properties of 2D materials is crucial for advancing both fundamental studies and technological applications.

In this context, standing-wave photoemission electron microscopy (SW-PEEM) has recently emerged as a powerful technique for depth-resolved spectromicroscopy, combining high spatial resolution with Ångstrom-scale depth sensitivity. The technique was first demonstrated by Kronast *et al.* [11] and later extended by Gray *et al.* [12] to study the depth-dependent magnetic and electronic properties of nanostructured materials. This approach relies on the generation of a soft X-ray standing wave within and above the sample by Bragg reflection from a multilayer mirror substrate [13-15]. By tuning the photon energy around the Bragg condition, the standing wave can be translated vertically through the sample, thereby enabling depth-resolved spectroscopy at distinct depths with sub-nanometer precision. Despite its demonstrated potential in nanostructured and magnetic materials, SW-PEEM has not yet been widely applied to the study of 2D materials and their heterostructures, even though precise depth sensitivity is particularly important in these systems.

In this study, we apply SW-PEEM to investigate the depth-dependent electronic structure and chemical bonding in monolayers of transition-metal dichalcogenides (TMDs), focusing on $WS_2$ and the influence of sulfur-related surface species on its near-surface electronic structure. By using a W/C multilayer mirror as a substrate, we generate a soft X-ray standing wave in the first-order Bragg reflection geometry, enabling depth-resolved spectromicroscopy with Ångstrom-scale precision. This approach allows us to track the evolution of the valence-band electronic structure and enhance sensitivity to contributions associated with the top and bottom sulfur layers within the $WS_2$ monolayer. Furthermore, we obtain an Ångstrom-resolved chemical depth profile of the sample by analyzing the standing-wave-induced modulations in core-level photoemission, thereby linking the electronic-structure results to the underlying chemical depth distribution. Experimental results are complemented by X-ray optical modeling and compared with first-principles density functional theory (DFT) calculations, including modeling of sulfur-related surface species, offering a detailed three-dimensional view of the electronic and chemical structure of monolayer $WS_2$.

This work establishes SW-PEEM as a powerful, non-destructive technique for depth-resolved imaging and spectroscopy of 2D materials, providing new insight into their electronic and chemical heterogeneity. The ability to probe valence-band states with depth sensitivity within a monolayer opens new avenues for exploring interfacial phenomena in van der Waals heterostructures and twisted 2D moiré systems, where depth-dependent interactions play a crucial role in emergent quantum phenomena. Future applications of this technique could provide critical insights into proximity effects, interlayer coupling, and electronic reconstruction at buried interfaces, thus paving the way for the rational design of next-generation 2D electronic and quantum devices.

Monolayer $WS_2$ was synthesized by chemical vapor deposition (CVD) on a growth substrate, retrieved using a polycarbonate/polydimethylsiloxane (PC/PDMS) stamp, and then transferred onto a W/C multilayer mirror substrate [16]. An optical image of the transferred flakes is shown in Figure 1a; the relatively low contrast between the flakes and the substrate arises from the high reflectivity of the multilayer substrate. Tapping-mode atomic force microscopy (AFM) yields a step height of ~8 Å, consistent with monolayer $WS_2$ (inset of Figure 1b). The optical quality of the transferred monolayer is further supported by room-temperature photoluminescence (PL), measured under ambient conditions with 532 nm excitation focused to a submicron spot, revealing a strong peak at ~1.93 eV and a weak shoulder at ~1.88 eV (Figure 1c). The dominant PL peak is consistent with monolayer $WS_2$, while the exciton redshift and weak shoulder likely reflect defect-induced localized states associated with local strain, substrate interactions, and other disorder-related effects [17-20]. Finally, micro-XPS confirms the expected W and S chemical signatures of the $WS_2$ monolayer (Figure 1d), further verifying the quality of the transferred sample.

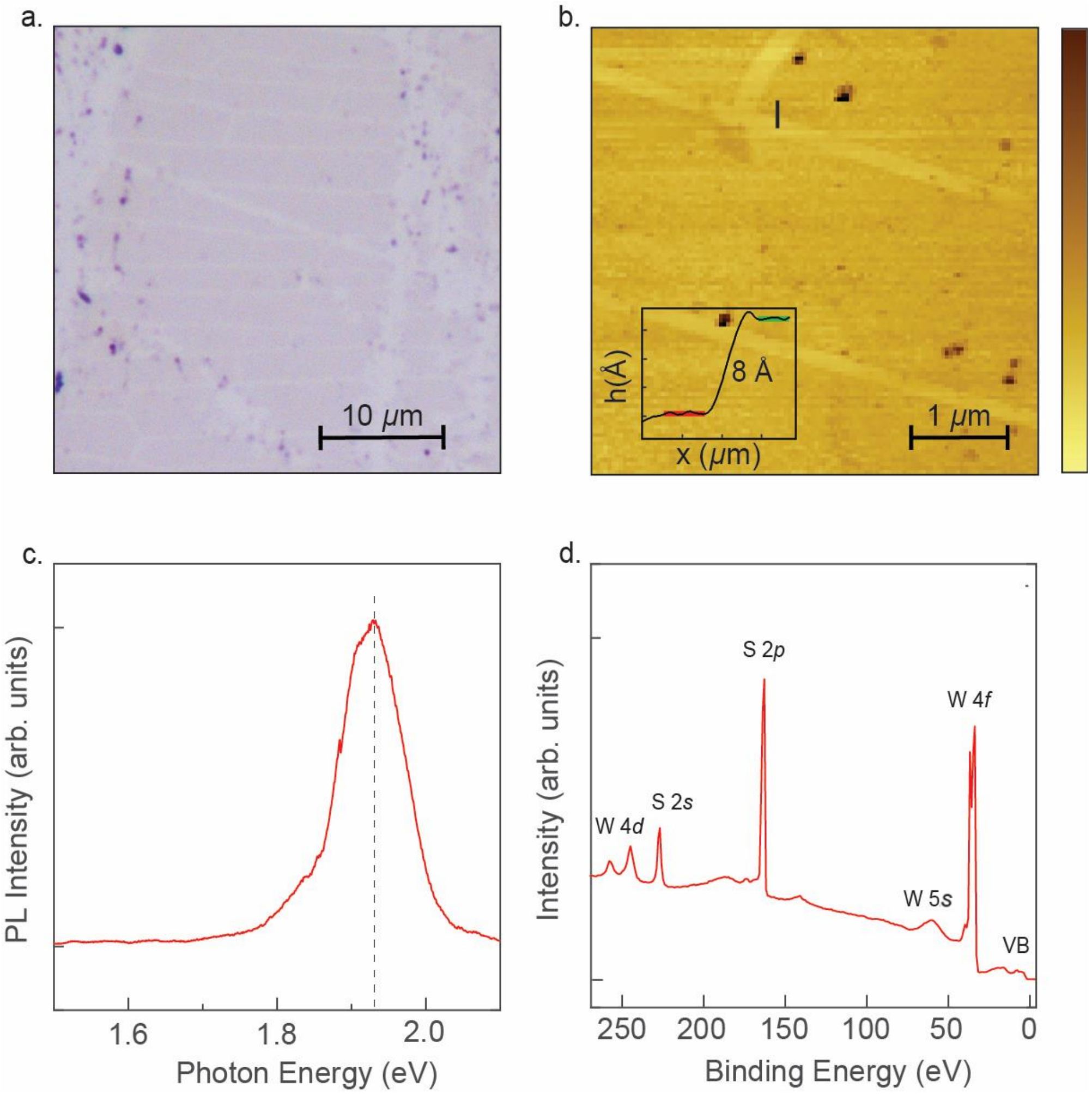

**Figure 1 | Structural, optical, and chemical characterization of monolayer $WS_2$ transferred onto a W/C multilayer mirror substrate.** (a) Optical micrograph of transferred $WS_2$ flakes on the multilayer substrate. (b) Tapping-mode AFM image with the height profile taken along the solid black line; the measured step height of ~8 Å is consistent with monolayer $WS_2$. (c) Room-temperature photoluminescence spectrum showing a dominant emission peak at ~1.93 eV and a weak shoulder at ~1.88 eV, consistent with monolayer $WS_2$ and indicative of disorder-related effects. (d) Micro-XPS core-level spectrum confirming the expected W and S chemical signatures of the $WS_2$ monolayer.

To investigate the depth- and spatially resolved electronic structure of the transferred $WS_2$ monolayer on the W/C multilayer mirror substrate, we employed kinetic-energy-filtered X-ray photoemission microscopy (XPEEM) at the SPEEM station of beamline UE49-PGM-a at the BESSY II storage ring (Helmholtz-Zentrum Berlin). The instrument is based on an Elmitec III photoemission electron microscope equipped with an integrated electrostatic photoelectron energy analyzer positioned between the objective column and the projective lenses. Acting as a kinetic-energy filter, this analyzer enables spectromicroscopic imaging using both core-level and valence-band photoelectrons. This capability is essential for the standing-wave measurements described below, as it enables depth-resolved spectromicroscopy of the electronic structure and chemical bonding. The instrument provides a typical lateral resolution of 30 nm.

For standing-wave depth-resolved measurements, SW-PEEM was performed in a standard first-order Bragg reflection geometry, schematically illustrated in Figure 2a. All data were acquired in normal-emission geometry, with the X-ray incidence angle fixed at 16°. The photon energy was scanned from 580 to 720 eV in 5 eV steps to vertically translate the standing wave generated by the W/C multilayer substrate through the $WS_2$ monolayer. The multilayer period of 3.6 nm was selected to satisfy the first-order Bragg condition under the chosen experimental geometry. This energy range spans the first-order Bragg peak of the W/C multilayer, where the standing-wave contrast is maximized, while remaining clear of the absorption edges of the constituent elements.

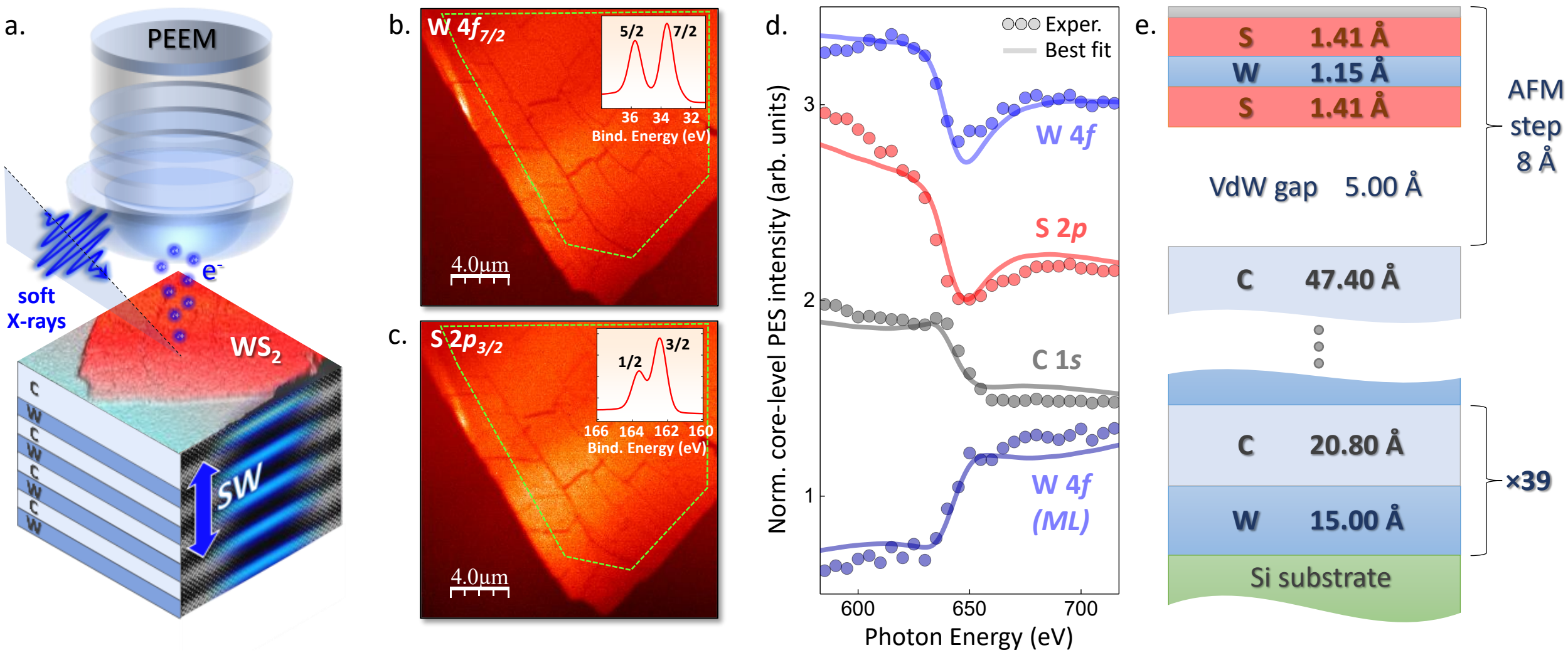

**Figure 2 | SW-PEEM imaging and structural modeling of a $WS_2$ monolayer on a W/C multilayer substrate.** (a) Schematic illustration of the standing-wave photoemission electron microscopy experiment performed in first-order Bragg reflection geometry and normal-emission detection. (b) Element-specific energy-filtered PEEM image of the $WS_2$ monolayer obtained from the W 4*f* core level, with the corresponding integrated photoelectron intensity spectrum extracted from the region of interest outlined in green shown in the inset. (c) Element-specific energy-filtered PEEM image of the $WS_2$ monolayer obtained from the S 2*p* core level, with the corresponding integrated photoelectron intensity spectrum extracted from the same region of interest shown in the inset. (d) Photon-energy-dependent photoemission yield curves of the W 4*f* and S 2*p* core levels from the $WS_2$ monolayer, together with the C 1*s* and substrate W 4*f* signals from the underlying W/C multilayer. Circular markers denote experimental data, and solid lines show the best-fit X-ray optical simulations. (e) Best-fit structural model of the sample obtained from simultaneous fitting of all measured yield curves shown in (d).

At each 5-eV photon-energy step of the standing-wave scan, a series of energy-filtered PEEM images was acquired while scanning the analyzer across the strongest accessible core-level peaks of the $WS_2$ monolayer, namely W 4*f* and S 2*p*. Element-specific energy-filtered PEEM images of the flake are shown in Figures 2b-c, with the corresponding integrated photoelectron intensity spectra displayed in the insets. The ability of PEEM to extract spectra from a defined region of interest (outlined in green) ensures that the substrate W 4*f* signal does not interfere with the W 4*f* signal originating from the $WS_2$ monolayer. In addition, the thickness of the amorphous

C capping layer (~4.7 nm) on the W/C multilayer substrate was optimized to suppress the substrate W 4*f* signal in the soft X-ray energy range, rendering it negligible compared to the signal from the $WS_2$ flake.

Spectra extracted from the defined regions of interest (either the $WS_2$ flake or the uncovered substrate) were first normalized to the storage ring current ($I_0$), corrected by subtraction of a Shirley-type background, and subsequently fitted with Voigt functions to determine the element-specific photoemission intensities. This procedure was applied consistently at each 5-eV photon-energy step of the standing-wave scan. The resulting intensities, plotted as circular markers in Figure 2d, are shown as a function of photon energy, which directly determines the vertical position of the standing wave within the sample. As previously demonstrated [12,14,15], these photoemission yield curves encode detailed information about the vertical dimensions of the individual layers in the structure and therefore form the basis for the structural modeling described below.

In addition to the W 4*f* and S 2*p* photoemission yield curves from the $WS_2$ monolayer flake, we extracted the C 1*s* signal, which originates primarily from the amorphous carbon capping layer on the W/C multilayer substrate. Furthermore, the weaker W 4*f* signal from the underlying W/C multilayer was obtained from the region of interest defined on the uncovered substrate. These additional data provide important constraints on the structural model, particularly for characterizing the substrate and determining the relative vertical position of the $WS_2$ flake with respect to the substrate.

To determine the vertical structure of the sample, we modeled the photon-energy-dependent photoemission intensities of the W 4*f*, S 2*p*, and C 1*s* core levels using the YXRO X-ray optical simulation software [21], including separate W 4*f* contributions from the flake and the substrate. The simulation calculates the X-ray standing-wave electric-field distribution within the sample and predicts the corresponding photoemission intensities as functions of photon energy, incidence angle (16°), and detection geometry, while accounting for subshell photoelectric cross sections and inelastic mean free paths. A single, self-consistent structural model was obtained by simultaneously fitting all measured yield curves using an automated optimization routine [22]. The final model is shown in Figure 2e, and the corresponding best-fit simulated intensities are overlaid as solid lines in Figure 2d, in excellent agreement with the experimental data in both peak positions and overall curve shapes.

The fitted structural parameters are physically plausible and consistent with independent measurements and prior reports. In particular, the measured W/C superlattice period (35.8 Å) is consistent with the nominal value of 36 Å, as calibrated by X-ray diffraction during deposition. The van der Waals gap is estimated to be 5 Å, and the total thickness of the $WS_2$ monolayer is found to be 3.97 Å. Thus, the total distance between the substrate surface and the top of the $WS_2$ layer is approximately 9 Å, in good agreement with the AFM-measured step height of the monolayer (~8 Å). Notably, this value is also consistent with previous reports of 6-10 Å, depending on sample type, measurement conditions, and preparation procedures such as annealing [23-25]. The observed van der Waals gap of 5 Å further agrees with experimentally determined interlayer distances between TMDC layers and other materials or substrates, such as h-BN, which are typically reported to be 5–6 Å in scanning transmission electron microscopy (STEM) measurements [26,27]. By contrast, theoretically predicted values for the van der Waals gap are typically smaller (~3-4 Å, depending on the specific layer combinations). This discrepancy is often attributed to adsorbed impurities at the interface, as well as to structural effects such as waviness or other forms of flexible deformation in the monolayer, both of which can increase the experimentally measured average gap [26].

The total $WS_2$ monolayer thickness obtained from the fit, 3.97 Å, compares well with prior STEM-based studies of TMDCs such as $MoS_2$, $WSe_2$, and $WS_2$, which typically report interatomic distances between the top and bottom chalcogen (S or Se) atoms in the range of 3.5-3.75 Å [26,27]. The slight quantitative discrepancy between these values and ours likely reflects differences in the measurement techniques. In cross-sectional STEM, particularly high-angle annular dark-field (HAADF) STEM, each bright spot corresponds to the projected intensity of electron scattering from a column of atoms, primarily determined by atomic number ($Z$) and column density. In contrast, standing-wave (SW) photoemission spectroscopy probes the depth distribution of photoemitting states (core or valence) and is therefore sensitive to the orbital character and hybridization of the contributing states. This distinction also helps explain why the individual thicknesses of the S atomic planes determined from SW-PEEM measurements of the S 2*p* core levels (1.41 Å) are greater than that of the W layer (1.15 Å), as measured using the W 4*f* core levels. This difference reflects the significantly larger radial extent of the sulfur 2*p* orbital compared to the more localized 4*f* orbital in tungsten. These findings highlight the potential of SW-based techniques for orbital-selective depth profiling in 2D materials and heterostructures and offer new insight into bonding and orbital hybridization at the atomic scale.

The X-ray optical model and the best-fit sample structure were used to calculate the depth-resolved profile of the X-ray standing-wave electric-field intensity ($E^2$) inside and above the sample. Figure 3a shows the experimentally derived structure of the uppermost layers together with the simulated standing-wave electric-field intensity as a function of depth and incident photon energy. The $E^2$ simulation indicates that, at lower photon energies (≈ 580-650 eV), the maximum intensity arises from an antinode located mainly beneath the $WS_2$ monolayer, predominantly in the van der Waals gap and the topmost C layer of the substrate. The first yellow dashed line marks a photon energy of 630 eV, at which the antinode just reaches the bottom S atomic plane, while the intensity at the height of the top S plane is strongly suppressed. The corresponding line profile of $E^2$ versus depth at 630 eV is plotted as the red curve in Figure 3b. By contrast, the second yellow dashed line marks 680 eV, at which the standing-wave field selectively enhances the signal from the topmost S atomic layer. Here, the second antinode lies largely above the sample and 'touches' the topmost S atoms. The resulting $E^2$ line profile is shown as the blue curve in Figure 3b.

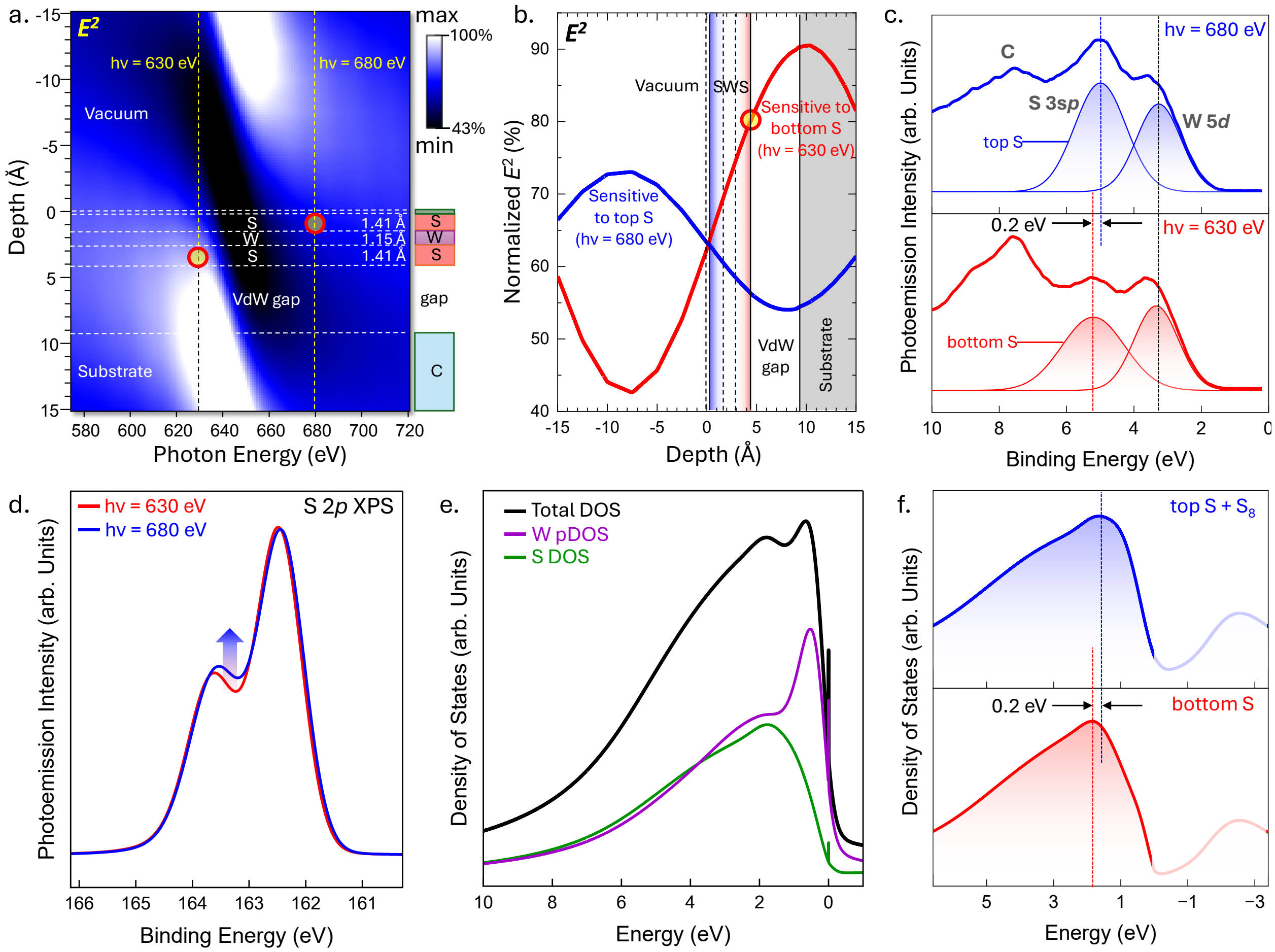

**Figure 3 | Standing-wave depth selectivity, valence-band response, and sulfur-related surface-species modeling in monolayer $WS_2$.** (a) Experimentally derived near-surface structure together with the simulated standing-wave electric-field intensity, $E^2$, as a function of depth and incident photon energy. (b) $E^2$ line profiles versus depth at 630 eV (red) and 680 eV (blue), corresponding to enhanced sensitivity to the bottom and top sulfur layers, respectively. (c) Valence-band spectra of monolayer $WS_2$ on the W/C multilayer recorded under standing-wave conditions at 680 eV (blue, top) and 630 eV (red, bottom). (d) S 2*p* core-level spectra recorded at 680 eV (more surface-sensitive) and 630 eV (more bulk-sensitive), showing enhanced spectral weight near ~163.5 eV in the more surface-sensitive spectrum. (e) Calculated density of states for the $WS_2$+$S_8$ system, decomposed into total (black), W 5*d* (magenta), and total S (green) contributions; the sulfur projection includes both $WS_2$ sulfur layers and the $S_8$ species. (f) Calculated sulfur-projected spectral weight with and without $S_8$, showing a ~0.2 eV shift toward the Fermi level upon inclusion of $S_8$, consistent in direction and approximate magnitude with experiment.

For quantitative comparison, the $E^2$ intensities shown in Figures 3a-b are normalized to the maximum intensity of the upper antinode, defined as 100%. A closer inspection of the red $E^2$ profile in Figure 3b reveals that the intensity at the depth of the top S atomic layer is 63.5%, whereas at the bottom S atomic layer it reaches 80%, corresponding to a ~26% enhancement in signal. This level of standing-wave contrast is comparable to, and in some cases exceeds, that reported in other standing-wave studies, including those of complex-oxide superlattices [28-30]. Thus, although the standing-wave period in the soft X-ray regime is significantly larger than the interatomic spacing in $WS_2$, appropriately tuned experimental conditions still enable selective enhancement of the photoemission signal from either the top or bottom S atomic layer by exploiting the spatial modulation of different antinodes.

The demonstrated depth sensitivity can also be extended to probe the valence-band electronic structure of $WS_2$ with depth resolution, thereby enhancing sensitivity to differences in the matrix-element-weighted densities of states (MEW-DOS) associated with the top and bottom sulfur layers within the monolayer. Figure 3c shows valence-band spectra of a $WS_2$ monolayer on the W/C multilayer, recorded under standing-wave conditions at photon energies of 680 eV (blue, top panel) and 630 eV (red, bottom panel). Both spectra display prominent peaks at ≈ 3.3 eV and ≈ 5.2 eV, arising predominantly from hybridized W 5*d* and S 3*sp* states, respectively, as well as a strong feature at ≈ 7.5 eV associated with C 2*sp* states in the underlying amorphous carbon layer.

The relative intensities of these valence-band features vary markedly between the two photon energies, revealing clear depth-dependent modulation of the valence-band response. These variations cannot be explained by small energy-dependent changes in photoionization cross sections [31] or inelastic mean-free paths [32] over this photon-energy range. Likewise, $k_z$-dependent band-structure effects and matrix-element variations can be ruled out because $WS_2$ is two-dimensional, the carbon layer is amorphous, and the spectra were collected over a wide analyzer acceptance angle that averages over a broad momentum range. Auger contributions in the 0–10 eV binding-energy window can also be dismissed at these non-resonant excitation energies. We therefore attribute the observed intensity modulation primarily to X-ray standing-wave effects near the first-order Bragg condition, in combination with the depth distribution of W, S, and C atoms in the heterostructure and the sensitivity of the standing wave to this depth distribution. This is clearly illustrated by the enhancement of the C 2*sp* feature (at ≈ 7.5 eV binding energy) in the red spectrum, measured at 630 eV, where the standing-wave antinode is positioned primarily within the van der Waals gap and the topmost carbon layer of the substrate, thereby enhancing the C signal (see Figs. 3a-b).

In addition to the depth-dependent modulation of valence-band intensity, the standing-wave measurements also reveal a relative binding-energy shift of approximately 0.2 eV in the spectral feature associated with strongly hybridized S 3*sp* states. To clarify the microscopic origin of this shift, we examine two physically motivated mechanisms: electronic coupling between the $WS_2$ monolayer and the underlying W/C multilayer substrate, and sulfur-related surface species that may alter the local electronic structure of the top sulfur layer.

To explain the experimentally observed ~0.2 eV shift between the top- and bottom-sulfur contributions to the S-derived valence band, we first considered the possible electronic interaction between $WS_2$ and the underlying W/C multilayer mirror substrate. Details of the calculations are presented in the Supporting Information together with Refs. [33-42]. DFT calculations using a 4 × 4 graphene supercell together with a 3 × 3 $WS_2$ supercell, where the graphene serves as a model of the carbon-terminated substrate surface, indicate that significant hybridization emerges only when the interlayer separation is artificially reduced to ~2.62 Å. The supercell structures are shown in Figure S1 of the Supporting Information, and the corresponding results are presented in Figure S2. This scenario, however, can be ruled out because both the experimentally determined van der Waals gap (5 Å) and the theoretically predicted value (3.42 Å) are significantly larger than 2.62 Å. Strong substrate-induced electronic coupling is therefore unlikely to account for the observed shift.

Having ruled out strong substrate-induced coupling, we next examined whether sulfur-related surface species could account for the observed spectral shift, taking advantage of the chemical sensitivity of micro-XPS. Specifically, we compared S 2*p* core-level spectra recorded at 680 eV, corresponding to a more surface-sensitive standing-wave geometry, and at 630 eV, corresponding to a more bulk-sensitive condition. A clear enhancement near ~163.5 eV is observed in the more surface-sensitive spectrum (Figure 3d). S 2*p* spectral weight in the 163.6–164.2 eV range is typically attributed to elemental sulfur species [43-45]. Because $S_8$ is the most common molecular form of elemental sulfur, this extra spectral weight is consistent with the possible presence of elemental-sulfur-like surface species, for which $S_8$ provides a physically motivated model.

To investigate this possibility, we modeled sulfur-related surface species using the energetically stable $S_8$ allotrope as a representative elemental-sulfur model and calculated the corresponding electronic structure. The theoretical spectra were then weighted by photon-energy-dependent photoionization cross sections and broadened using Gaussian and Lorentzian functions to account for instrumental resolution and lifetime effects, respectively, thereby enabling direct comparison with experiment (see Figs. 3e-f). Figure 3e shows the calculated density of states for the $WS_2$+$S_8$ system, decomposed into total (black), W 5*d* (magenta), and total S (green) projections; the latter includes contributions from both the $WS_2$ sulfur layers and the $S_8$ species. The W 5*d* states dominate the spectral weight near the Fermi level, consistent with the prominent low-binding-energy W 5*d* feature observed in both experimental spectra (Figure 3c). Inclusion of $S_8$ enhances the sulfur-projected density of states and produces a shift of ~0.2 eV toward the Fermi level, capturing the direction and approximate magnitude of the experimentally observed shift within the calculational uncertainty (see Figure 3f). These results indicate that sulfur-related surface species contribute to the layer-dependent electronic structure and likely play an important role in the observed energy shift in $WS_2$. Details of the calculations are presented in the Supporting Information document, and the corresponding supercell is shown in Figure S3.

In summary, we demonstrate standing-wave photoemission electron microscopy as a powerful platform for depth-resolved spectromicroscopy of two-dimensional van der Waals materials using monolayer $WS_2$ on a W/C multilayer mirror substrate. By translating the soft X-ray standing-wave field through the monolayer, we achieve Ångstrom-scale sensitivity to its vertical electronic and chemical structure, including depth-dependent modulation of the valence-band response and a chemically resolved depth profile. The measurements reveal enhanced sensitivity to states associated with the top and bottom sulfur layers despite the atomic-scale thickness of the monolayer. Combined X-ray optical modeling and first-principles calculations further indicate that the observed ~0.2 eV sulfur-derived valence-band shift is unlikely to arise from strong direct substrate-induced hybridization and is instead consistent with contributions from sulfur-related surface species. These results establish SW-PEEM as a non-destructive probe of depth-dependent electronic structure in 2D materials and highlight its potential for studying interfacial coupling, chemical reconstruction, and emergent states in van der Waals heterostructures and moiré systems.

**Supporting Information**

Additional laboratory- and synchrotron-based characterization, X-ray optical modeling, and DFT methods and results (PDF).

**Notes:**

The authors declare no competing financial interest.

Correspondence and requests for materials should be addressed to A.X.G.

## ACKNOWLEDGMENTS

J.R.P., S.S., and A.X.G. acknowledge support from the U.S. Department of Energy, Office of Science, Office of Basic Energy Sciences, Materials Sciences and Engineering Division under Award No. DE-SC0024132. A.X.G. also gratefully acknowledges the support from the Alexander von Humboldt Foundation. R.M. and J.K. acknowledge support from the U.S. Department of Energy, Office of Science, Office of Basic Energy Sciences, under Award Nos. DE-SC0020323 and DE-SC0025490. J.K. also acknowledges graduate student support for A.T. through NSF CAREER Award No. DMR-2339309. The work at Northeastern University was supported by the National Science Foundation through the Expand-QISE award NSF-OMA-2329067 and benefited from the resources of Northeastern University's Advanced Scientific Computation Center, the Explorer Cluster, the Massachusetts Technology Collaborative award MTC-22032, and the Quantum Materials and Sensing Institute (QMSI). The authors thank the Helmholtz-Zentrum Berlin für Materialien und Energie for allocation of synchrotron radiation beamtime on beamline UE49-PGM-a at the BESSY II electron storage ring (212-10264-ST).

# *Supporting Information*

# Imaging of van der Waals Materials via Standing-Wave Photoemission Microscopy: Depth-Resolved Electronic Structure of $WS_2$

Jay R. Paudel[1,2], Ryan Muzzio[3], Matthew E. Matzelle[4,5], Sharup Sheikh[1], Ahmed Fadul[4,5], Aalok Tiwari[3], Farhad Salmassi[6], Eric Gullikson[6], Kathleen M. McCreary[7], Berend T. Jonker[7], Jouko Nieminen[8,4], Claus M. Schneider[9], Florian Kronast[10], Arun Bansil[4,5], Jyoti Katoch[3], and Alexander X. Gray[1]

[1] *Physics Department, Temple University, Philadelphia, Pennsylvania 19122, USA*

[2] *Chemical Sciences Division, Lawrence Berkeley National Laboratory, Berkeley, California 94720, USA*

[3] *Department of Physics, Carnegie Mellon University, Pittsburgh, Pennsylvania 15213 USA*

[4] *Department of Physics, Northeastern University, Boston, Massachusetts 02115, USA*

[5] *Quantum Materials and Sensing Institute, Northeastern University, Burlington, Massachusetts 01803, USA*

[6] *Center for X-ray Optics, Lawrence Berkeley National Laboratory, Berkeley, California 94720, USA*

[7] *U.S. Naval Research Laboratory, Washington, District of Columbia 20375, USA*

[8] *Computational Physics Laboratory, Tampere University, Tampere 33014, Finland*

[9] *Peter Grünberg Institut (PGI-6), Forschungszentrum Jülich GmbH, D-52425 Jülich, Germany*

[10] *Helmholtz-Zentrum Berlin für Materialien und Energie, 12489 Berlin, Germany*

**email: axgray@temple.edu*

**Computational Details**

We performed density functional theory (DFT) calculations using projector-augmented-wave (PAW) pseudopotentials [1], as implemented in the Vienna ab initio simulation package (VASP) [2-5]. The Perdew-Burke-Ernzerhof (PBE) exchange-correlation functional [6] was used, and van der Waals (vdW) corrections were included using Grimme's DFT-D3 method [7]. A plane-wave cutoff energy of 520 eV was employed. Any method-specific choices are outlined in the appropriate section.

**Figure S1: $WS_2$ on Graphene**

To determine whether the asymmetry observed in the depth-resolved photoemission spectra of monolayer $WS_2$ arises from the influence of the W/C multilayer substrate on the bottom sulfur layer, we performed two-dimensional slab calculations using DFT with a 4×4 graphene substrate and a 3×3 $WS_2$ overlayer separated by 20 Å of vacuum (see Figure S1). The pseudopotentials used in this calculation had the following valence electron configurations: C ($2s^2$ $2p^2$), S ($3s^2$ $3p^4$), W ($6s^1$ $5d^5$). The calculations were self-consistent until the energy difference between successive electronic steps was less than 10 μeV. The lateral lattice constants of the supercell were set by graphene (a = b = 2.42 Å) [8], and the $WS_2$ overlayer was strained to match. Compared to the experimental lattice constants of $WS_2$ (a = b = 3.153 Å) [9], this corresponds to a compressive epitaxial strain of about 4%; however, this strain is not expected to strongly affect the interlayer coupling of interest.

**Figure S2: Interlayer-Distance Dependence of Sulfur DOS**

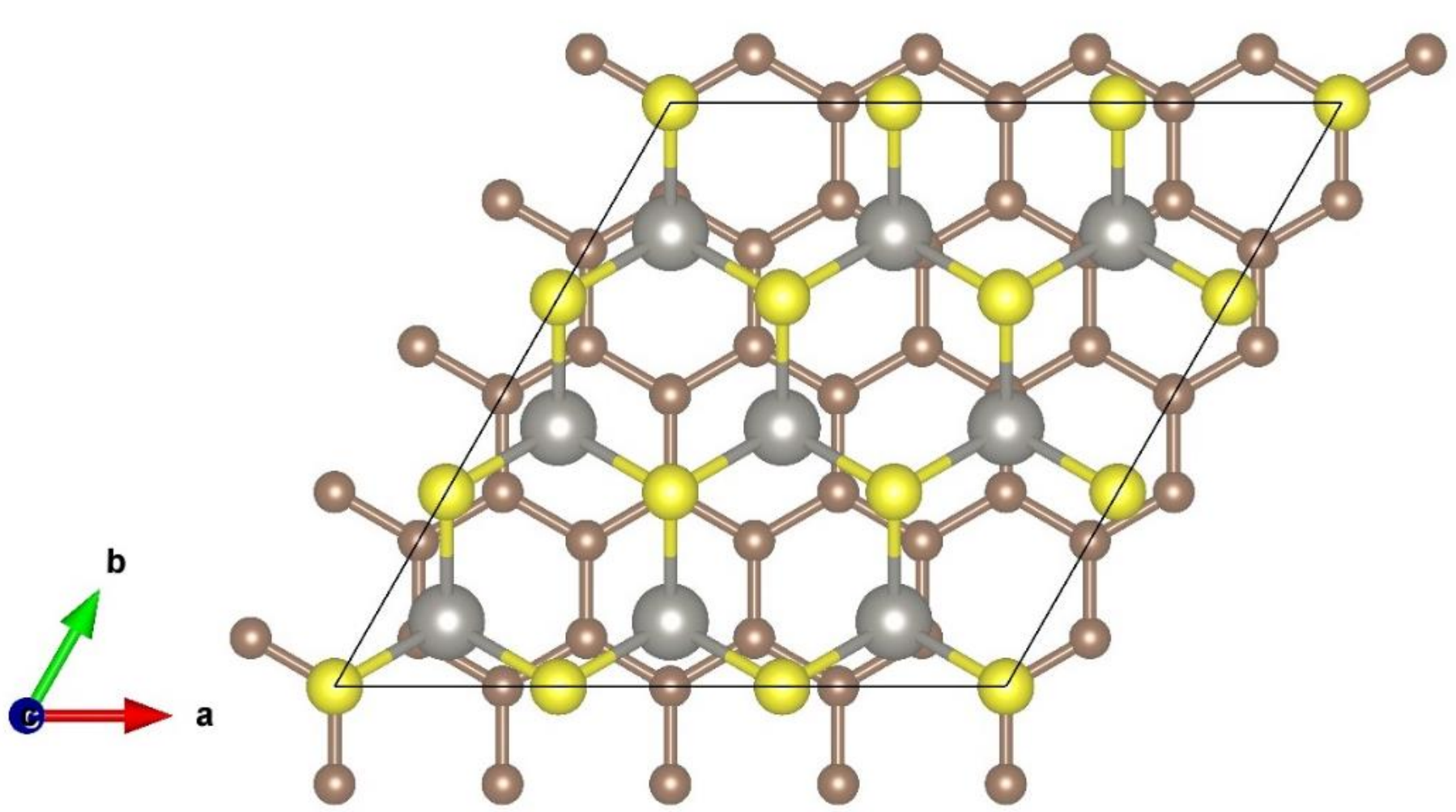

**Figure S1:** Supercell containing heterostructure of 4×4 graphene and 3×3 $WS_2$ looking down the *c*-axis. Carbon is depicted as brown spheres, tungsten as silver spheres, and sulfur as yellow spheres, and bonds are drawn between nearby atoms. The software VESTA [10] was used to generate the figure.

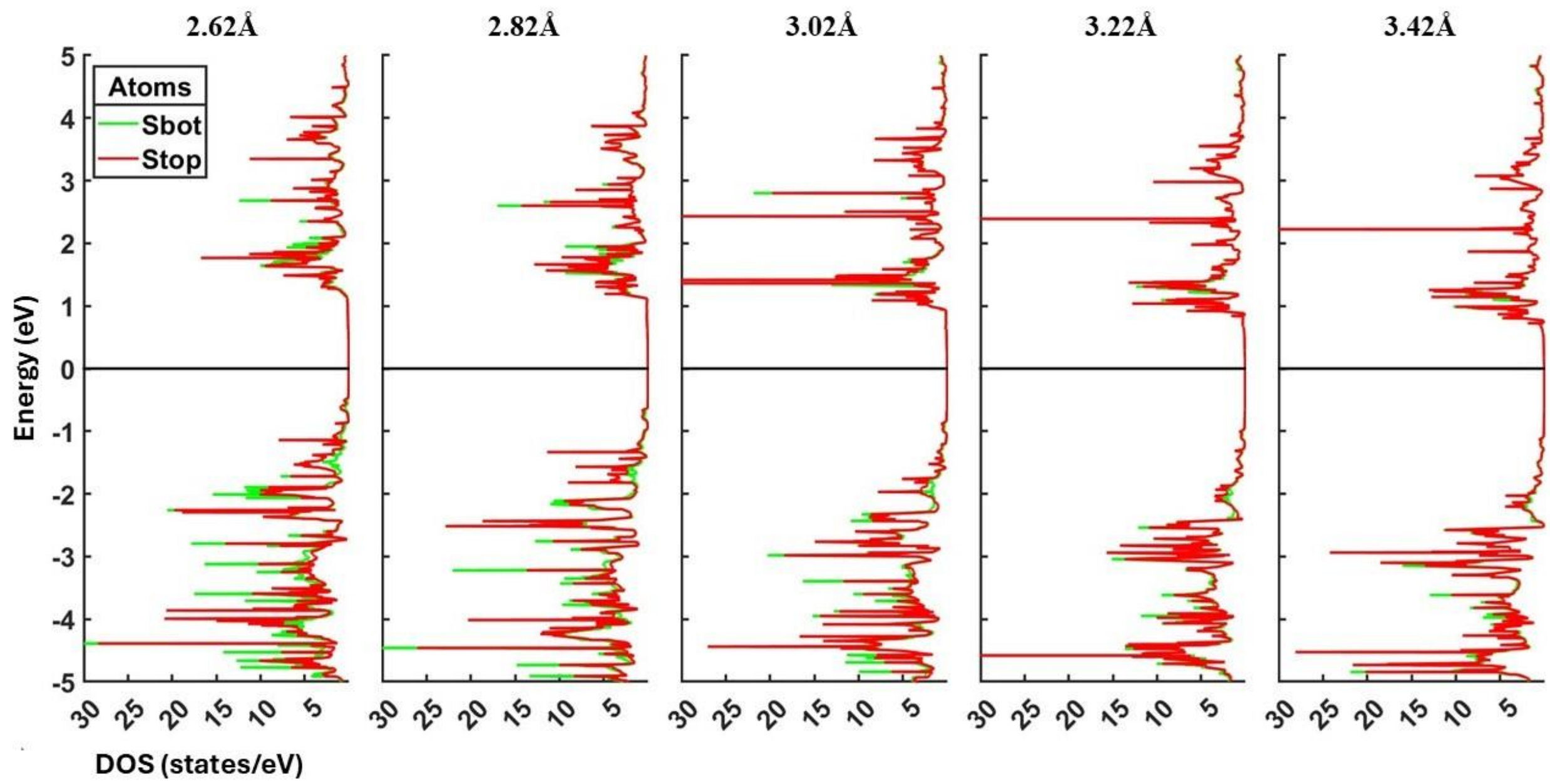

**Figure S2:** The element decomposed DOS for the top (red) and bottom (green) sulfur are shown in the near Fermi energy range. From left to right, the interlayer distance between the graphene and the $WS_2$ ranges from 2.62 Å to 3.42 Å, which is the calculated equilibrium distance. Appreciable differences between the top and bottom sulfur only manifest at the lowest interlayer distance.

To predict the equilibrium vertical spacing between the layers, atomic relaxation was performed using the conjugate gradient method, with the graphene atoms held fixed and the $WS_2$ atoms allowed to relax in all directions, until the atomic forces were reduced below 8 meV/Å. The equilibrium distance, measured from the plane of the bottom sulfur atoms to the plane of the graphene substrate, was found to be 3.42 Å, with minute variations in the heights of the bottom sulfur atoms of less than 0.001 Å. The density of states (DOS) near the Fermi level from this calculation is shown in Figure S2. The bottom- and top-sulfur contributions, plotted in red and green, respectively, show no appreciable difference that could account for the observed 0.2 eV shift in the experiment.

To determine whether a smaller interlayer distance would lead to an appreciable difference between the bottom- and top-sulfur density of states (DOS), we performed several calculations for interlayer distances ranging from 3.42 to 2.62 Å. The corresponding results are also shown in Figure S2. An appreciable difference between the bottom- and top-sulfur DOS emerges only at 2.62 Å.

However, this distance is much smaller than the measured vdW gap of 5 Å, making it highly unlikely that any region of the experimental heterostructure approaches such a separation. The observed 0.2 eV shift must therefore originate from a different source.

**Figure S3: $S_8$ on $WS_2$**

Another possible origin of the difference in the depth-resolved photoemission spectra is the presence of contaminants on the $WS_2$ surface. To test this possibility, we performed a two-dimensional slab calculation using DFT for a 5×5 $WS_2$ substrate with an $S_8$ molecule positioned 7.5 Å above it and with 20 Å of vacuum (see Figure S3). The pseudopotentials used in this calculation included the following valence electron configurations: S ($3s^2$ $3p^4$), W ($5s^2$ $5p^6$ $5d^6$), where the W GW version was used. The calculations were self-consistent until the energy difference between successive electronic steps was less than 1 μeV. A 5 × 5 $WS_2$ supercell was found to be large enough to eliminate interactions between $S_8$ molecules in adjacent unit cells. The equilibrium distance was found to be 2.7 ± 0.4 Å by rigidly shifting the $S_8$ molecule from 1.9 to 11.5 Å above the $WS_2$ layer in steps of 0.4 Å within the 5

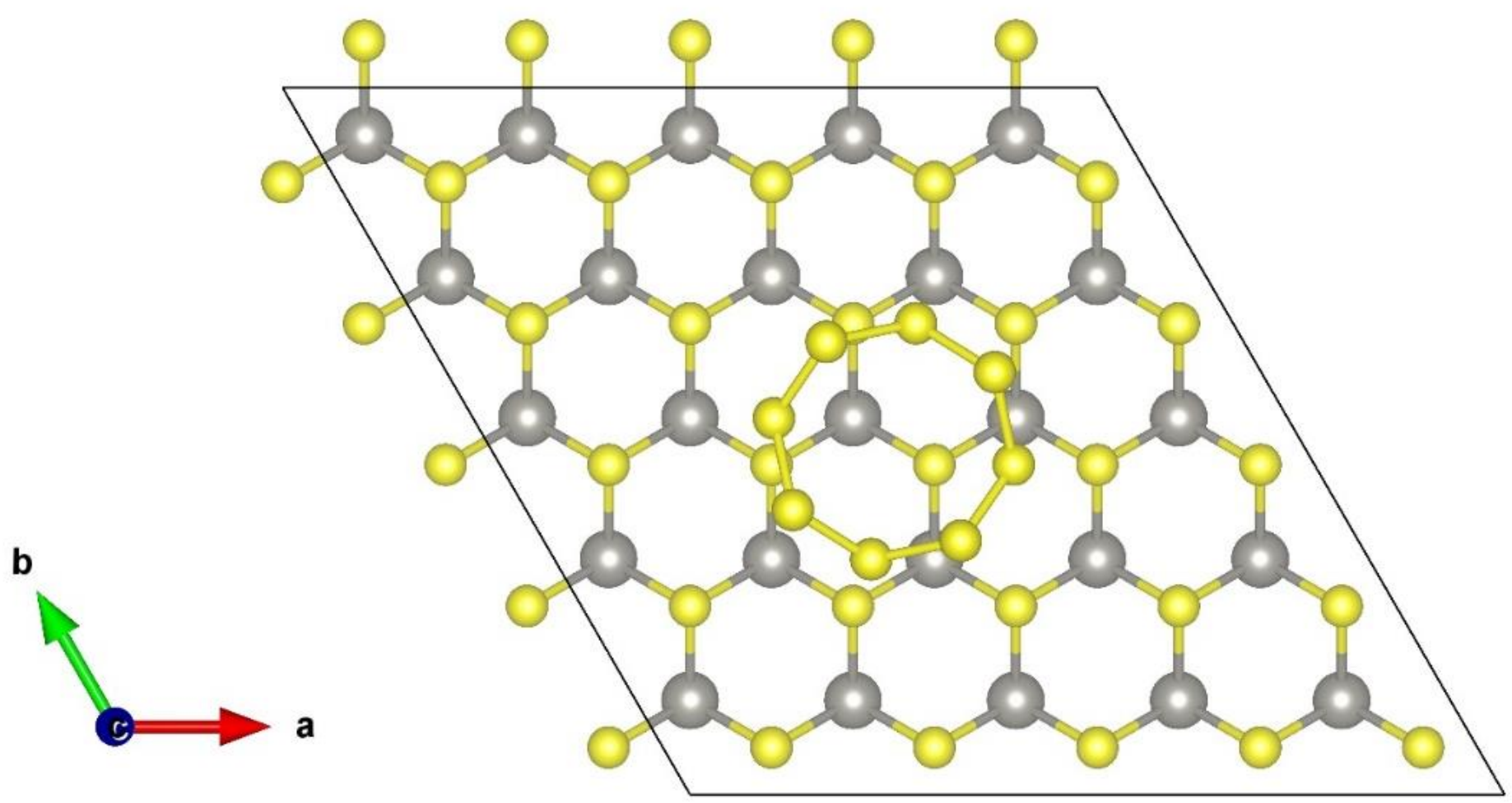

**Figure S3:** Supercell containing heterostructure of 5×5 $WS_2$ and $S_8$ molecule looking down the c-axis. Tungsten is depicted as silver spheres and sulfur as yellow spheres, and bonds are drawn between nearby atoms. The software VESTA [10] was used to generate the figure.

× 5 supercell. However, the calculation used for comparison with experiment was performed at an interlayer distance of 7.5 Å in order to eliminate bonding interactions between $WS_2$ and $S_8$. This choice avoids the additional complexity of determining the lowest-energy relative lateral positions of the molecule and substrate, which would be computationally costly.

In this limit, the $WS_2$ and $S_8$ are contained within the same simulation cell, but their electronic structures remain effectively noninteracting, so that the total density of states (DOS) is well approximated by the sum of the individual $WS_2$ and $S_8$ contributions. The sulfur-related contribution from $S_8$ was obtained from the projected density of states (PDOS) by summing the orbital contributions over all atoms in the molecule. Although the DOS associated with an isolated $S_8$ molecule is relatively discrete, the total calculated spectrum becomes smoother because it is combined with the $WS_2$ DOS and subsequently weighted by photon-energy-dependent photoionization cross sections and broadened with Gaussian and Lorentzian functions to account for instrumental resolution and lifetime effects, respectively. In this way, a noninteracting $WS_2$+$S_8$ spectrum is constructed and modified to enable direct comparison with the experimental spectra.